\documentclass[11pt]{iopart}
\usepackage{bm}
\usepackage{hyperref}
\usepackage{amssymb}
\usepackage{feynmp}
\usepackage{amsopn}
\usepackage{setstack}
\usepackage{multirow}
\usepackage{deltandiagrams}
\newcommand{\eg}{\emph{eg.}}

\renewcommand{\etal}{\emph{et al.}}
\newcommand{\uv}{\mathrm{UV}}
\newcommand{\eternal}{\mathrm{sr}}
\newcommand{\phiplanck}{\phi_{\mathrm{P}}}

\newcommand{\fnl}{f_{\mathrm{NL}}}
\newcommand{\nnl}{n_{\mathrm{NL}}}
\newcommand{\Ps}{\mathcal{P}}
\newcommand{\Planck}{M_{\mathrm{P}}}

\newcommand{\E}{\mathbb{E}}
\newcommand{\dS}{\mathrm{dS}}

\renewcommand{\d}{\mathrm{d}}
\newcommand{\vect}[1]{\bm{\mathrm{{#1}}}}
\renewcommand{\e}[1]{\mathrm{e}^{{#1}}}

\newcommand{\im}{\mathrm{i}}

\renewcommand{\geq}{\geqslant}

\DeclareMathOperator{\RePart}{Re}

\renewcommand{\Re}{\RePart}
\DeclareMathOperator{\Ei}{Ei}
\DeclareMathOperator{\Ci}{Ci}

\newenvironment{equation-diagram}{\vspace{3mm}\begin{equation}}
  {\end{equation}\par\vspace{-3mm}\noindent\ignorespaces\hspace{-0.7ex}}
\newenvironment{eqnarray-diagram}{\vspace{3mm}\begin{eqnarray}}
  {\end{eqnarray}\par\vspace{-3mm}\noindent\ignorespaces\hspace{-0.7ex}}

\makeatletter
\newcommand\numberwithin[2]{\@addtoreset{#1}{#2}}
\makeatother
\numberwithin{footnote}{section}

\begin{document}
\begin{fmffile}{diags}
	\title{One-loop corrections to the curvature perturbation from inflation}
	\date{\today}
	\author{David Seery}
	\address{Astronomy Unit, School of Mathematical Sciences\\
	  Queen Mary, University of London\\
	  Mile End Road, London E1 4NS\\
	  United Kingdom}
	\eads{\mailto{D.Seery@qmul.ac.uk} and \mailto{djs61@cam.ac.uk}}
	\submitto{JCAP}
	\pacs{98.80.-k, 98.80.Cq, 11.10.Hi}
	\begin{abstract}
		An estimate of
		the one-loop correction to the power spectrum of the
		primordial curvature perturbation is given,
		assuming it is generated during a
		phase of single-field, slow-roll inflation.
		The loop correction splits into two parts, which can
		be calculated separately: a purely
		quantum-mechanical contribution which is generated from the
		interference among quantized field modes around the time
		when they cross the horizon, and a classical contribution
		which comes from
		integrating the effect of field modes which have already
		passed far beyond the horizon.
		The loop correction contains logarithms which may
		invalidate the use of na\"{\i}ve perturbation theory
		for cosmic microwave background (CMB) predictions
		when the scale associated with the CMB is exponentially
		different from the scale at which the fundamental theory
		governing inflation is formulated.
	\vspace{3mm}
	\begin{flushleft}
		\textbf{Keywords}:
		Inflation,
		Cosmological perturbation theory,
		Physics of the early universe,
		Quantum field theory in curved spacetime.
	\end{flushleft}
	\end{abstract}
	\maketitle

	\section{Introduction}
	For some years, the theory of inflation
	\cite{Starobinsky:1980te,Sato:1980yn,Guth:1980zm,
	Hawking:1981fz,Albrecht:1982wi,Linde:1981mu,Linde:1983gd}
	has represented our most
	successful approach to the very early universe.
	Although we do not know with certainty what physical conditions applied
	in the very distant past, it is believed that a number
	of fundamental properties of our universe were determined at
	that time.
	Among the most important of these is the ensemble
	of small density perturbations which later condensed into galaxies
	and gave rise to the network of cosmic structure we see today.
	Therefore, one question which must
	be answered by any theory which purports to
	describe the very early universe concerns the origin and nature of these
	perturbations.
	In inflationary models, the answer is provided by use of an era of
	quasi-exponential expansion to generate small
	fluctuations from the vacuum.
	Remarkably, the properties of these perturbations
	can be calculated by applying a minimal extension of quantum field theory
	to the expanding universe
	\cite{Bardeen:1983qw,Guth:1982ec,Hawking:1982cz,Hawking:1982my,
	Mukhanov:1985rz,Sasaki:1986hm,Mukhanov:1989rq,Mukhanov:1990me}.
	
	The inflationary perturbations are generated by quantum-mechanical
	effects among field modes of wavelength $k$ around the time when such
	modes are of order the Hubble scale, that is, $k \sim aH$.
	If the inflationary expansion is still ongoing when these
	perturbations are generated then the local Hubble scale continues to
	contract, so that eventually they are far outside the horizon,
	giving $k \ll aH$. In this regime
	the field behaves in an approximately classical fashion
	\cite{Lyth:2006qz}. Therefore
	the evolution of the curvature perturbation can
	be calculated reliably until the end of inflation,
	when it is generally assumed to settle down to a time-independent
	final value, unless isocurvature modes exist which drive
	its evolution on large scales
	\cite{Lyth:2001nq,Moroi:2001ct,Mollerach:1989hu}.
	In this simple picture we have a robust prediction of the
	properties of the perturbations regardless of the details of the
	inflationary era.

	The above account neglects at least one possible source of trouble,
	namely the question of quantum mechanical corrections.
	In any quantum theory, including inflation,
	we are obliged to study the role of so-called \emph{loop corrections},
	which may modify or destroy the leading semiclassical prediction.
	Loop corrections are systematic corrections in powers of
	a coupling constant, which is small if perturbation theory is to be a
	good approximation.
	However, perturbation series are generally asymptotic rather than
	convergent, and if the coefficients of higher-order terms
	are large then such corrections need not negligible even if the
	coupling constant is small.
	In such circumstances, these corrections must be taken
	into account when performing accurate comparisons of theory with
	observational data; for example, their use is routine when interpreting
	the outcome of particle physics experiments.
	Modern cosmological datasets have become remarkably precise
	\cite{Spergel:2006hy,Martin:2006rs,Kinney:2006qm},
	and observational improvements
	imply that future datasets will possess even greater constraining power.
	It is important that our theoretical
	predictions keep pace with observational developments.
	For this reason,
	as in particle physics, we should ask whether loop
	corrections can modify the predictions we use to compare with experiment.
	
	If such corrections are to play a role, what form should we expect them
	to take?
	Models of inflation make predictions concerning the properties of the
	primordial curvature perturbation, $\zeta$. This perturbation is
	communicated to the cosmic microwave background (CMB) at
	decoupling.
	The statistical properties of $\zeta$ are
	predicted by supposing that the universe is geometrically
	close to a patch of the classical, eternal
	de Sitter universe, in which the
	expansion is exactly exponential.	
	The de Sitter universe has a special
	simplicity as a solution of the Einstein equations since it is maximally
	symmetric, implying that its curvature is constant. However,
	as an arena for physics the de Sitter universe is considerably less simple
	\cite{Witten:2001kn}.
	A horizon surrounds any freely
	falling observer and screens a portion of the spacetime from view.
	This horizon is associated with a finite entropy,
	$S_{\dS}$, which can be thought of
	as representing a finite state space of size $\sim \exp S_{\dS}$
	available to each inertial observer. One therefore expects that perturbative
	predictions made over a volume $V \gtrsim \exp S_{\dS}$
	(in Planck units) are untrustworthy,
	since the volume $V$ carries more perturbative degrees of freedom
	than can actually be available in the full quantum gravity
	\cite{ArkaniHamed:2007ky,Giddings:2007nu}.
	An inflationary epoch lasting for
	$N$ e-folds causes the universe to expand by a factor $\e{N}$,
	so we expect that the inflationary era cannot be described in
	perturbation theory for an arbitrarily
	long time; in order not to violate the volume bound, it is necessary
	that $N \lesssim S_{\dS}$ for reliable calculations.
	Related limits on $N$ of various sorts have been reported in the
	literature \cite{Wu:2006ew,Losic:2006ht,Huang:2007zt}.

	The above intuitive argument implies that we should expect the
	coefficients in the loop expansion to become large when $N \gg 1$.
	For sufficiently large $N$, we will have a breakdown of perturbation
	theory and thus an apparent loss of predictivity from the early universe.
	The purpose of the present paper is to explore the details of
	this loss of predictivity, quantified as
	a loop correction to the observable perturbation
	$\zeta$. Although a proper understanding of the breakdown of perturbation
	theory is important as a matter of principle, it is also
	important in practice. The apparent success of the inflationary
	prediction means that it has
	become routine for experiments which measure
	temperature anisotropies in the cosmic microwave background
	to report their findings in terms of constraints on inflationary
	models. These constraints are usually derived using tree-level predictions
	for $\zeta$ which ignore the possibility of a breakdown in perturbation
	theory.

	The important ramifications which attach to a failure of the perturbative
	description mean that
	loop corrections have received attention from many authors,
	beginning with 't Hooft \& Veltman \cite{Hooft:1974bx}.
	The possibility of an infrared perturbative breakdown was explored
	by Sasaki, Suzuki, Yamamoto \& Yokoyama
	\cite{Sasaki:1992ux,Suzuki:1992gi},
	and later authors computed loop corrections in a variety of different
	settings;
	for example, Refs.~%
	\cite{Mukhanov:1996ak,Abramo:1997hu,Abramo:1998hi,Unruh:1998ic,
	Prokopec:2002uw,Onemli:2002hr,Prokopec:2003bx,Onemli:2004mb,
	Brunier:2004sb,
	Boyanovsky:2005sh,Boyanovsky:2005px,Kahya:2005kj,Weinberg:2005vy,
	Chaicherdsakul:2006ui,Losic:2005vg,Kahya:2006hc,Bilandzic:2007nb,
	Kahya:2006ui,Prokopec:2006ue,Sloth:2006az,Sloth:2006nu,
	Kim:2007sb}.
	We will derive an estimate for the loop correction
	for an arbitrary potential,
	using the slow-roll approximation to control the calculation.
	A central feature of this analysis is the use of
	the non-linear $\delta N$ expansion
	\cite{Starobinsky:1986fx,Sasaki:1995aw,Lyth:2004gb,Lyth:2005fi}.
	This expansion provides a
	systematic method for computing the evolution of $\zeta$
	(and its correlation functions) in terms of the field
	perturbations at earlier times (and their correlation functions).
	Moreover, as will be described below,
	the $\delta N$ formulation has many useful properties
	for the purposes of loop calculations.
	In order to apply this method consistently, it is necessary
	to account for the presence of loop corrections in the field
	perturbations at early times. The requisite
	loop correction has recently been
	calculated \cite{Seery:2007we}, and in this paper all these
	corrections are assembled into a loop correction for
	$\zeta$, which allows an estimate of the onset of any breakdown
	in perturbation theory
	in a number of popular models of inflation.
	
	This paper is organized as follows.
	In \S\ref{sec:deltaN} the non-linear $\delta N$ formula is
	briefly reviewed, and in \S\ref{sec:oneloop} it is used to derive
	the complete one-loop correction to the power spectrum of
	$\zeta$. Our result agrees with that of Byrnes \etal,
	who recently gave an expression for the loop correction valid to
	two loops. A feature of the $\delta N$ expansion is that one is
	free to construct the initial hypersurface at any time after the relevant
	$\vect{k}$-modes have left the horizon. This freedom is
	discussed in \S\ref{sec:slice}.
	In \S\ref{sec:rg-power} we use the formal $\delta N$
	expression to compute an estimate for the loop correction.
	The loop correction is shown to be divergent on large
	scales in \S\ref{sec:rg-log}, and must be regularized by restricting
	the computation to some finite box of size $\ell$.
	In \S\S\ref{sec:constant-tilt}--\ref{sec:monomials} estimates for
	the regularized loop correction are given in two approximations.
	In \S\ref{sec:constant-tilt} the spectrum is assumed to
	be defined by a constant tilt, whereas in
	\S\ref{sec:monomials} we specialise to the case of a monomial potential
	and compute an explicit, but model-dependent, estimate.
	Finally, we conclude with a discussion of these results
	in \S\ref{sec:conclude}.
	
	We use natural units in which the speed of light $c$,
	Planck's constant $\hbar$ and the Planck mass
	$\Planck \equiv (8\pi G)^{-1/2}$ are set equal to unity.
	The metric signature is $(-,+,+,+)$ and the unperturbed background
	metric is taken to be of Friedmann--Robertson--Walker form
	\begin{equation}
		\d s^2 = - \d t^2 + a(t)^2 \d \vect{x}\cdot \d \vect{x} ,
		\label{eq:background-metric}
	\end{equation}
	where $t$ is cosmic time and $a$ is the scale factor of the universe.
	It is often more convenient to
	use a conformal time variable, $\eta$, defined by
	$\eta \equiv \int_{\infty}^{t} \d t / a(t)$.
	The Hubble parameter satisfies $H \equiv \dot{a}/a$, where
	an overdot denotes a derivative with respect to $t$, and
	inflation occurs whenever $\epsilon \equiv - \dot{H}/H^2 < 1$.

	Our theory will consist of Einstein gravity coupled to a
	scalar field $\phi$ with potential $V(\phi$), which can be
	arbitrary except that it must support an epoch of inflation for
	some values of the field. The background is taken to be
	homogeneous with spatially-dependent perturbations $\delta\phi$
	which satisfy the smallness condition $|\delta\phi| \ll |\phi|$.
	When $\epsilon \ll 1$, the field $\phi$ rolls only a short distance in
	a Hubble time, and therefore this is known as the slow-roll
	regime. When slow-roll applies, it is often a good formal
	approximation to compute in powers of $\epsilon$ or related small
	quantities obtained by taking dimensionless time derivatives of
	$\epsilon$, although a key theme of this paper will be that such
	expansions are usually asymptotic rather than convergent.
	The most important related parameter is $\eta$,
	not to be confused with the conformal time,
	and defined by $\eta \equiv \ddot{H}/H\dot{\phi}$.
	This measures the local curvature in the potential $V$,
	and in typical models it is small whenever $\epsilon$ is, although
	this need not be true in general.
	
	\section{The $\delta N$ formula}
	\label{sec:deltaN}
	
	Our theories which govern cosmological evolution at high
	energies are typically written in terms
	of a number of microscopic fields whose quanta are the particle
	species which populate the universe.
	The interesting cosmological observables are
	expectation values of products of these fields, often in the special
	case where the fields which participate in the expectation value
	are evaluated at equal times but at distinct spatial positions
	$\{ \vect{x}_1, \ldots, \vect{x}_n \}$.
	It is usually more convenient to translate to Fourier space,
	and work instead with expectation values of fields evaluated at
	distinct wavenumbers $\{ \vect{k}_1, \ldots, \vect{k}_n \}$.
		
	\paragraph{The power spectrum and bispectrum.}
	\label{sec:spectra}
	The lowest-order expectation value is the one-point function,
	\begin{equation}
		\langle \delta\phi(\vect{k}) \rangle = (2\pi)^3
		\delta(\vect{k}) O ,
		\label{eq:onepoint}
	\end{equation}
	where $O$ is a dimensionless quantity which depends on the time
	of evaluation. At tree-level the one-point function is always
	zero if we have chosen to work in a perturbatively stable vacuum,
	but it is possible that a non-zero $O$ may
	be generated radiatively. It was shown in Ref.~\cite{Seery:2007we}
	that to one-loop order
	$O$ is given by a renormalization-scheme dependent number,
	which can be absorbed into a redefinition of the background field.

	The next-order expectation value is the two-point function,
	\begin{equation}
		\langle \delta\phi(\vect{k}_1) \delta\phi(\vect{k}_2) \rangle
		= (2\pi)^3 \delta(\vect{k}_1 + \vect{k}_2) P(k_1) ,
		\label{eq:twopoint}
	\end{equation}
	where $P(k)$ again depends on the time of evaluation and is
	known as the power spectrum. It is sometimes more convenient
	to work in terms of the
	``dimensionless'' power spectrum $\Ps$, which is
	related to $P$ by the rule $\Ps \equiv k^3 P/2\pi^2$.
	
	The three-point expectation value can be written
	\begin{equation}
		\langle \delta\phi(\vect{k}_1)\delta\phi(\vect{k}_2)
		\delta\phi(\vect{k}_3) \rangle = (2\pi)^3
		\delta(\vect{k}_1+\vect{k}_2+\vect{k}_3)
		B(k_1,k_2,k_3) ,
		\label{eq:threepoint}
	\end{equation}
	where $B$ is referred to as the \emph{bispectrum}, since in
	typical cases it is proportional to a sum of terms quadratic in $P$
	\cite{Bartolo:2004if,Komatsu:2001rj}.
	To express $B$ in terms of a dimensionless quantity, it is
	conventional to introduce the momentum-dependent parameter $\fnl$
	\cite{Komatsu:2001rj},
	analogous to $\Ps$ and occasionally called the ``reduced bispectrum,''
	which is defined to satisfy
	\cite{Maldacena:2002vr,Lyth:2005fi}%
		\footnote{Sometimes $\fnl$ is defined by writing
		$\zeta = \zeta_g - \frac{3}{5} \fnl \star \zeta_g^2$
		(in coordinate space), where $\zeta$ is the full curvature
		perturbation, $\zeta_g$ is a gaussian random field,
		`$\star$' denotes a convolution
		and $\fnl$ is to be thought of as parametrizing a possible
		$\chi^2$-type contribution to $\zeta$. To linear order in $\fnl$,
		this expansion for $\zeta$ in terms of $\zeta_g$ produces the
		formula~\eref{eq:fnl-def}. However, since there is no motivation
		for such a specific type of non-linearity in $\zeta$, we follow
		Lyth \& Rodr\'{\i}guez \cite{Lyth:2005fi}
		in adopting Eq.~\eref{eq:fnl-def}
		for $\fnl$, whatever its source.}
	\begin{equation}
		B(k_1,k_2,k_3) \equiv - \frac{6}{5} \fnl \Big\{
			P(k_1)P(k_2) + P(k_2)P(k_3) + P(k_3)P(k_1)
		\Big\} .
		\label{eq:fnl-def}
	\end{equation}
	Note that $B$ (and $\fnl$)
	can always be written as a function of the three
	scalars $\{ k_1, k_2, k_3 \}$, in terms of which it does not depend
	directly on the relative orientation of the $\{ \vect{k}_i \}$;
	alternatively, one can use the momentum conservation
	condition $\sum_i \vect{k}_i = 0$ to rewrite $B$
	as a function of only two of the momenta, but in this case
	it will explicitly depend on their relative orientation, making $B$
	(and $\fnl$)
	always a function of three independent degrees of freedom.
	
	\paragraph{Tilt.}
	The dimensionless quantities $\Ps$ and $\fnl$ are generally
	scale-dependent. Their scale dependence can locally be characterized
	by two \emph{tilt} parameters $n$ and $\nnl$ \cite{Chen:2005fe}, defined by
	\begin{equation}
		n - 1 \equiv \frac{\d \ln \Ps}{\d \ln k}
		\quad \mbox{and} \quad
		\nnl \equiv \frac{\d \ln \fnl}{\d \ln k} ,
		\label{eq:runnings}
	\end{equation}
	where in the definition of $\nnl$
	we take $\fnl$ to be evaluated at equilateral momenta
	$k_i = k$, and
	by convention $n-1$ is chosen as the tilt of $\Ps$ rather
	than $n$.%
		\footnote{Note that $\nnl$ clearly does not capture the most general
		variation in $\fnl$ under a change of the momenta, since
		$\fnl$ is a function of three degrees of freedom.
		More generally, one could define the tilt of $\fnl$ as a vector
		quantity obtained by taking the gradient with respect to the
		$k_i$.}
	These numbers are typically small \cite{Kosowsky:1995aa}.
	For a limited range of
	wavenumbers it may be an acceptable approximation to
	take $\Ps$ and $\fnl$ as almost flat over the range, whereas if it
	is necessary to characterize $\Ps$ or $\fnl$ over a larger
	range then one can write
	\begin{equation}
		\Ps \approx \left.\Ps\right|_{k_0}
		\left( \frac{k}{k_0} \right)^{(n-1)|_{k_0}}
		\quad \mbox{and} \quad
		\fnl \approx \left.\fnl\right|_{k_0}
		\left( \frac{k}{k_0} \right)^{\left.\nnl\right|_{k_0}}
		\label{eq:running-approximations}
	\end{equation}
	where $k_0$ is some pivot wavenumber which is characteristic of
	the range of wavenumbers in question, and `$|_{k_0}$' denotes
	evaluation at this point. However, both these approximations
	break down over an exponentially large range of wavenumbers since
	in general none of
	$\Ps$, $\fnl$, $n$ or $\nnl$ are constant. Over such an
	exponentially large range
	Eq.~\eref{eq:running-approximations} substantially overpredicts
	the spectrum, which typically grows only like a power of
	$\ln k$ rather than a power of $k$
	\cite{Weinberg:2005vy,Weinberg:2006ac}. Where possible we will use
	Eqs.~\eref{eq:runnings}--\eref{eq:running-approximations} to
	give model-independent estimates for tilted spectra,
	but it is necessary to make more detailed
	model-dependent estimates
	where an exponentially large range of scales applies
	(see \S\ref{sec:monomials}).
	
	\paragraph{The $\delta N$ formula.}
	The microscopic fields are not observable by themselves; only a
	specific combination of them can be observed as the gauge-invariant
	perturbation $\zeta$ \cite{Bardeen:1980kt}
	which is communicated to the CMB. An extremely powerful
	and computationally convenient way to obtain this combination is
	to employ the so-called $\delta N$ formula
	\cite{Starobinsky:1986fx,Sasaki:1995aw,Rigopoulos:2004gr,
	Lyth:2004gb,Lyth:2005fi}.
	Consider the Arnowitt--Deser--Misner
	\cite{Arnowitt:1960es} line element
	\begin{equation}
		\d s^2 = - N \, \d t^2 + h_{ij}
		(
			\d x^i + N^i \, \d t
		)
		(
			\d x^j + N^j \, \d t
		) ,
	\end{equation}
	in which the perturbations are parametrized
	by a lapse function $N$, the shift vector $N^i$ and the spatial
	metric $h_{ij}$.
	For any choice of $t$, we can write the spatial metric on the
	three-dimensional slices of constant time in the form
	$h_{ij} = a(t)^2 \exp\{ 2\psi(t,\vect{x}) \} \delta_{ij}$.%
		\footnote{This is not strictly true, because this is not the
		most general perturbation of a spatial slice. The curvature
		perturbation $\psi$ amounts to a fluctuation in the trace
		of $h_{ij}$, which describes the volume expansion in a local
		region of the slice. In principle there are further
		vector and tensor modes which describe vorticity and gravitational
		waves. These modes are set to zero.
		In general this is \emph{not} a consistent procedure,
		because scalar, vector and tensor perturbations mix beyond
		linear order. Therefore the presence of scalar perturbations
		will induce perturbations in the vector and tensor modes.
		However, we expect that these effects will not destroy
		the leading order prediction for $\zeta$, which we can obtain by
		focusing simply on $\psi$.}
	The quantity $\psi$ is known as the \emph{curvature perturbation}
	associated with this slicing.
	
	Consider any two hypersurfaces at times $t_{\ast}$
	(the \emph{initial} hypersurface) and
	$t_c$ (the \emph{final} hypersurface), where $t_c > t_\ast$.
	In the perturbed universe the
	number of e-foldings of inflation between these slices
	will vary from place to place, and satisfies
	\begin{equation}
		N(t_\ast \rightarrow t_c, \vect{x}) = N_0(t_\ast \rightarrow t_c) +
		\psi_c(\vect{x}) - \psi_\ast(\vect{x}) ,
	\end{equation}
	where $N_0$ represents the number of e-folds which would have taken
	place in the unperturbed background. By choosing the initial slice to
	be flat, so that $\psi_\ast = 0$, and the final slice to be uniform
	density, so that $\psi_c = \zeta_c$, one obtains the simple expression
	$\zeta = \delta N$, where $\delta N \equiv N(\vect{x}) - N_0$
	is the shift in e-folds between $t_\ast$ and $t_c$ generated by the
	perturbations.
	We expect widely separated Hubble volumes to evolve locally like
	independent universes, so on large scales this shift can be computed
	merely by
	accounting for the variation in initial conditions from place to
	place at time $t_\ast$,
	\begin{equation}
		\zeta_c(\vect{x}) = \sum_{n=1}^\infty
			\left\{ \delta\phi_{\ast}(\vect{x}) \right\}^n
			\Big( \frac{\partial}{\partial \phi_\ast} \Big)^n
			N(t_c,t_\ast) ,
			\label{eq:deltaN}
	\end{equation}
	where the $\delta\phi_\ast$ are the field perturbations at
	$t_\ast$, evaluated on the flat slicing. (Note that $\zeta_c$
	defined by this expression has no dependence on $t_\ast$.)
	In principle $N$
	depends on all the initial conditions on the initial slice,
	and because $\phi$ is governed by a second-order differential equation
	it will generally be necessary to specify the pair
	$\{ \phi_\ast,\dot{\phi}_\ast \}$ at $t_\ast$.
	We will make the simplifying assumption that the
	slow-roll approximation applies at the initial time.
	This relates
	$\dot{\phi}_\ast$ to $\phi_\ast$, and hence
	it is unnecessary to include $\dot{\phi}_\ast$ separately.
	
	\paragraph{Expectation values.}
	Eq.~\eref{eq:deltaN} can be used to compute $n$-point
	correlation functions of the
	curvature perturbation $\zeta$, by first constructing a product
	of $n$ copies
	of~\eref{eq:deltaN} and taking expectation values
	$\langle \cdots \rangle$ in the
	appropriate vacuum state. After application of this process,
	and working to lowest non-trivial order in $\delta\phi$,
	it follows that
	\begin{equation}
		\langle \zeta(\vect{x}_1) \zeta(\vect{x}_2) \rangle_c =
		\Big\{
			\frac{\partial N(t_c,t_\ast)}{\partial \phi_\ast}
		\Big\}^2
		\langle \delta\phi(\vect{x}_1) \delta\phi(\vect{x}_2)
		\rangle_\ast ,
		\label{eq:zeta-tree-spectrum}
	\end{equation}
	where the subscript $c$ denotes evaluation on the final slice
	at time $t_c$.
	Eq.~\eref{eq:zeta-tree-spectrum} shows that we can compute
	the power spectrum at any late time $t_c$
	using our knowledge of correlators only at some fixed early
	time $t_\ast$. By repeating the above process for
	higher $n$-point expectation values and to higher orders
	in $\delta\phi$, it is obvious that we can use the
	$\delta N$ formula to generate predictions for any
	$n$-point function of $\zeta$ to any required accuracy
	compatible with the neglect of spatial gradients.
	
	The manipulations described above are now standard in the literature.
	However, the foregoing discussion makes it completely clear that
	the $\delta N$ formula is merely an
	identity which expresses $\zeta$ in terms of geometrical quantities.
	From this point of view, $\zeta$ is nothing more than
	a book-keeping object
	which describes how much one region of the universe has expanded
	(if $\zeta > 0$) or contracted (if $\zeta < 0$) relative to the
	mean expansion. Therefore, as has been observed before,
	it is a pure gauge mode in exact de Sitter space \cite{Maldacena:2002vr}.
	
	How are we to assign an interpretation to the notion of mean expansion?
	This is necessary to determine what question we actually answer
	when we compute properties of $\zeta$ using the $\delta N$ formula.
	It corresponds to specifying which region of the universe is supposed
	to be described by the unperturbed background
	metric~\eref{eq:background-metric}. In practice, we do not imagine that
	the inflationary patch which gave rise to the large flat region we
	observe today fills our entire spatial slice. For example,
	this may happen because the universe emerged from some primeval
	phase with
	initial conditions which were chaotically varying, and the requirements for
	successful inflation were only realised at a fraction of spatial
	positions.
	Arguments of this sort are invoked for chaotic inflationary models, and
	arise in a similar way for cosmologies based on the so-called
	``landscape'' of string theory vacua.
	In such scenarios we should expect spacetime to retain a
	disordered, inhomogeneous and anisotropic character beyond the boundary
	of the inflationary patch.
	
	It is quite clear that we should only associate the FRW background
	metric~\eref{eq:background-metric} with the region of spacetime
	containing the primordial inflationary patch,
	perhaps after a few e-foldings of inflation so that the patch is
	sufficiently close to spatial flatness and isotropy
	\cite{Wald:1983ky}.
	One then expects that whatever primordial unpleasantness lurks beyond
	the boundary is irrelevant when we make predictions
	from inflation, provided we live sufficiently far from the boundary.
	This is analogous to the decoupling theorem in field theory
	(see, {\eg}, \cite{Collins:1984xc}).
	It follows from the simple point of principle, related to
	cluster decomposition, that in order to make a prediction at one point
	we should not need to know the disposition of the universe at
	some far distant location.
	
	In practice this means that $a(t)$ should be interpreted as
	providing only an effective description within some box
	which coincides with the inflationary patch.
	But we are quite free, if we wish, to imagine that we are working within
	some smaller box, as long as it is large enough to contain scales of
	order the present horizon size.
	Whichever box size we pick, $\zeta$ represents
	the modulation in local expansion as one varies one's location within
	the box. Do the answers to calculations
	depend on the box size? Certainly.
	In fact, $a(t)$ will generally shift in
	value between differently sized boxes
	\cite{Lyth:1991ub,Boubekeur:2005fj,Lyth:2006gd,Lyth:2007jh}, because the
	split between background and perturbations is different on disparate
	scales.
	One can obtain the correct variation in $a(t)$
	by demanding that the expectation value of $\zeta$
	vanish within the box, that is, $\E_{\ell}(\zeta) = 0$, where
	$\E_{\ell}$ represents the spatial expectation value within a box
	of characteristic size $\ell$. This process, which determines the
	physical perturbation mode within the box, is analogous to imposing
	a renormalization condition which determines a physical property
	of a particle, such as its mass.
	
	This $\ell$-dependence of $a(t)$ and $\zeta$ imply that
	one is left with predictions for the properties of $\zeta$
	which depend on the box size, $\ell$.
	As emphasized by Boubekeur \& Lyth, one can aggregate predictions
	in boxes of size $\ell$ into larger boxes of size $\ell' \gg \ell$
	in such a way that the dependence of $\zeta$ on $\ell$ is
	cancelled by the variation of background quantities
	\cite{Boubekeur:2005fj,Lyth:2007jh}.
	This is analogous to renormalization group invariance in field theory.
	Nevertheless, the statistical properties of $\zeta$ itself
	depend on $\ell$.
	At first sight which seems like a failure of locality.
	The resolution is that if the properties of $\zeta$ calculated
	within a small box are accessible to experiment, the properties
	of $\zeta$ calculated some much larger box are not, because they require
	measurements to be made on scales which are larger than our present horizon.
	In order to make predictions for the CMB, we wish $\zeta$ to represent
	the perturbation which is actually present in the radiation when the
	universe becomes transparent. This entails making an optimal choice
	for $\ell$ in the same way that making a prediction for a
	scattering experiment at some characteristic momentum transfer
	entails making an optimal choice for the running coupling constant.

	In what follows we compute the loop correction in
	$\langle \zeta(\vect{k}_1) \zeta(\vect{k}_2) \rangle$. This will enable
	us to discuss the $\ell$-dependence of the prediction, and select
	an appropriate box size for the purposes of CMB predictions.
	
	\section{The one-loop correction to the power spectrum
	of $\zeta$}
	\label{sec:oneloop}
	
	\subsection{A renormalized $\delta N$ formula}
	\label{sec:renormalized-delta-n}
	
	In this section, we compute the formal $\delta N$ expression
	for the one-loop correction to $\langle \zeta(\vect{k}_1)
	\zeta(\vect{k}_2) \rangle$. This expression will turn
	out to be divergent and will require,
	as outlined in \S\ref{sec:deltaN} above.
	In order to organize the calculation, it is convenient to express
	the various contributions to $\zeta$ using a set of
	graphical rules, analogous to the Feynman diagrams of quantum
	field theory. Similar diagrams were introduced by
	Crocce \& Scoccimarro \cite{Crocce:2005xy} and adapted for use
	in the present context by Zaballa, Rodr\'{\i}guez \& Lyth
	\cite{Zaballa:2006pv}.
	The use of such diagrams has recently been formalized
	to all orders by Byrnes {\etal} \cite{Byrnes:2007tm} using a slightly
	different notation.
	Their principal use is to keep track of
	various loop corrections which arise from the $\delta N$
	algorithm described above; when Eq.~\eref{eq:deltaN} is translated to
	Fourier space, the products $\delta\phi(\vect{x})^n$ at a
	point $\vect{x}$ become convolutions with independent integrals
	over $n-1$ momenta. Some of these momenta are removed after
	taking expectation values, but this cancellation is not always
	exact and any unconstrained integrals can be thought of as
	making loop corrections to the expectation value.
	 
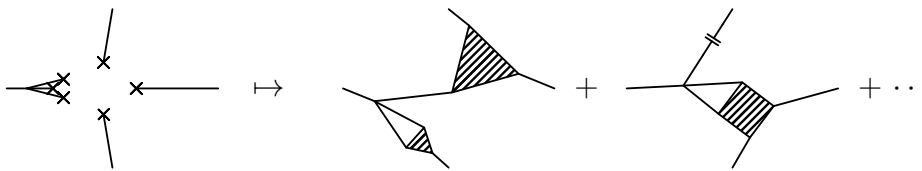
\begin{figure}
		\begin{center}
			\parbox{25mm}{\begin{fmfgraph*}(80,60)
				\deltaNdefs
				\fmfpen{0.8thin}
				\fmfleft{l}
				\fmfright{r}
				\fmftop{t}
				\fmfbottom{b}
				\fmf{plain,tension=5}{l,v1}
				\fmf{plain}{v1,t1}
				\fmf{plain}{v1,t2}
				\fmf{plain}{v1,t3}
				\fmf{plain}{b,t4}
				\fmf{plain}{r,t5}
				\fmf{plain}{t,t6}
				\fmfv{decoration.shape=cross,decoration.size=0.075w}{t1}
				\fmfv{decoration.shape=cross,decoration.size=0.075w}{t2}
				\fmfv{decoration.shape=cross,decoration.size=0.075w}{t3}
				\fmfv{decoration.shape=cross,decoration.size=0.075w}{t4}
				\fmfv{decoration.shape=cross,decoration.size=0.075w}{t5}
				\fmfv{decoration.shape=cross,decoration.size=0.075w}{t6}
				\fmf{phantom,tension=1.25}{t1,t2}
				\fmf{phantom,tension=1.25}{t2,t3}
				\fmf{phantom,tension=1.25}{t3,t4}
				\fmf{phantom,tension=1.25}{t4,t5}
				\fmf{phantom,tension=1.25}{t5,t6}
				\fmf{phantom,tension=1.25}{t6,t1}
			\end{fmfgraph*}}
			\hspace{5mm}
			$\mapsto$
			\hspace{5mm}
			\parbox{25mm}{\begin{fmfgraph*}(80,60)
				\deltaNdefs
				\fmfpen{0.8thin}
				\fmfleft{l}
				\fmfright{r}
				\fmftop{t}
				\fmfbottom{b}
				\fmf{plain,tension=5}{l,v1}
				\fmf{plain}{v1,t1}
				\fmf{plain}{v1,t2}
				\fmf{plain}{v1,t3}
				\fmf{plain,tension=5}{b,t4}
				\fmf{plain,tension=5}{r,t5}
				\fmf{plain,tension=5}{t,t6}
				\fmfpoly{shaded}{t2,t3,t4}
				\fmfpoly{shaded}{t1,t5,t6}
			\end{fmfgraph*}}
			\hspace{3mm}
			$+$
			\hspace{1mm}
			\parbox{25mm}{\begin{fmfgraph*}(80,60)
				\deltaNdefs
				\fmfpen{0.8thin}
				\fmfleft{l}
				\fmfright{r}
				\fmftop{t}
				\fmfbottom{b}
				\fmf{plain,tension=2}{l,v1}
				\fmf{plain}{v1,t1}
				\fmf{plain}{v1,t2}
				\fmf{plain}{v1,t3}
				\fmf{plain,tension=2}{b,t4}
				\fmf{plain,tension=2}{r,t5}
				\fmf{plain,tension=2}{t,t6}
				\fmf{powerspectrum_plain}{t1,t6}
				\fmfpoly{shaded}{t2,t3,t4,t5}
			\end{fmfgraph*}}
			\hspace{3mm}
			$+$ $\cdots$
		\end{center}
		\caption{Diagrammatic representation of the $\delta N$ expansion.
		To represent an $n$-point correlator one draws all $m$-valent
		vertices attached to
		$n$ external legs. Internal lines
		terminate at special vertices marked by a cross `$\times$'.
		One then draws all possible $j$-gons which connect the
		$\times$-vertices, of which two possible examples are displayed.
		2-gons are treated specially for
		notational convenience (see Fig.~\ref{fig:one-two-gons}).
		Each $m$-valent vertex contributes a ``coupling constant''
		$\partial^{m-1} N/\partial \phi_\ast^{m-1}$.
		\label{fig:diagrams}}
	\end{figure}

\begin{figure}
		\begin{center}
			\parbox{25mm}{\begin{fmfgraph*}(80,40)
				\deltaNdefs
				\fmfpen{thin}
				\fmfleft{l}
				\fmfright{r}
				\fmf{plain}{l,v1}
				\fmf{phantom}{v1,v2}
				\fmf{plain}{v2,r}
				\fmfv{decoration.shape=cross,decoration.size=0.075w}{v1}
				\fmfv{decoration.shape=cross,decoration.size=0.075w}{v2}
			\end{fmfgraph*}}
			\hspace{5mm}
			$\mapsto$
			\hspace{5mm}
			\parbox{25mm}{\begin{fmfgraph*}(80,40)
				\deltaNdefs
				\fmfpen{thin}
				\fmfleft{l}
				\fmfright{r}
				\fmf{powerspectrum_plain}{l,r}
			\end{fmfgraph*}}
		\end{center}
		\caption{Special representation for
		2-gon diagrams. In the left-hand diagram
		the two $\times$-vertices should be connected by a 2-gon, but this is
		difficult to draw. Instead, we adopt the notation shown in the
		right-hand diagram,
		where the 2-gon is represented by two parallel
		lines transverse to the propagators in which they are inserted.
		In principle there should also be 1-gon initial conditions, where
		a single $\times$-vertex is tied off at a one-point function
		$\langle \delta\phi \rangle$, but because we are neglecting
		the one-point function on large scales such 1-gons do not appear
		in these diagrams.
		\label{fig:one-two-gons}}
	\end{figure}
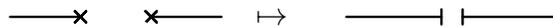

	The rules for drawing diagrams can be summarised as follows.
	In order to evaluate the contribution to an $n$-point correlator,
	one draws all diagrams with $n$ external legs which inject
	momenta $\{ \vect{k}_i \}$ into the diagram, where $i = 1, \ldots, n$.
	To each external
	leg one attaches a single
	$m$-valent vertex (where $m \geq 2$), at which momentum
	conservation applies. Each $m$-valent vertex contributes
	a ``coupling constant'' $\partial^{m-1} N/\partial \phi^{m-1}$.
	The internal lines attached to such
	vertices terminate at any number of special vertices,
	which we label with a cross `$\times$'. One then draws
	all combinations of $j$-gons which can be used to connect
	the $\times$-vertices with no $\times$-vertex left over, replacing
	each $j$-gon by a $j$-point expectation value of the
	$\delta\phi$ as shown in Fig.~\ref{fig:diagrams}.
	These expectation values should be
	evaluated at the momenta which flow into the
	corresponding $\times$ vertices.
	For clarity, we allow a special notation
	for	2-gons. (See Fig.~\ref{fig:one-two-gons}.) Finally one
	integrates over all internal momenta in order to obtain the
	expectation value. In such diagrams, it is sometimes useful
	to imagine the various $j$-gons as representing initial conditions
	which flow from the
	interior of the graph
	to the outside, encountering time evolution at
	each $m$-valent vertex.
	
	
	The discussion in \S\ref{sec:deltaN} implies that we
	should generalize the formal $\delta N$ expansion, Eq.~\eref{eq:deltaN},
	by inserting a zero-momentum expectation value $\delta N_0$
	\begin{equation}
		\zeta = \delta N_0 + \sum_{n=1}^\infty
			\Big\{ \delta\phi_{\ast}(\vect{x}) \Big\}^n
			\left( \frac{\partial}{\partial \phi_\ast} \right)^n
			N(t_c,t_\ast) ,
			\label{eq:deltaN-renormalized} ,
	\end{equation}
	which is to be adjusted so that $\E_\ell( \zeta )  = 0$, meaning that
	$\zeta$ accurately represents the perturbation in expansion
	over a box of characteristic size $\ell$.
	Enforcing this condition shows that $\delta N_0$ must obey
	\begin{equation}
		\delta N_0 = - \sum_{j = 2}^{\infty} \frac{1}{j!}
			\frac{\partial^j N}{\partial \phi^j}
			\langle \delta\phi^j \rangle ,
		\label{eq:deltaN-zero}
	\end{equation}
	where we have assumed that $\langle \delta \phi(\vect{x}) \rangle = 0$
	and that statistical homogeneity forces
	$\langle \delta\phi^j \rangle \equiv \langle \delta\phi^j(\vect{x})
	\rangle$ to be independent of $\vect{x}$ for $j \geq 2$.
	This guarantees that $\delta N_0$ carries zero momentum.
	Angle brackets $\langle \cdots \rangle$ denote an expectation value in the
	quantum vacuum at past infinity, as usual, and $\E_\ell$ denotes
	a spatial average over a comoving spherical box of characteristic size
	$\ell$.

	The scale factor is not the only background quantity which will
	receive a renormalization.
	In general we can expect all parameters which specify the homogeneous
	background model to each receive
	renormalizations.%
		\footnote{This is similar to the proposal of Kolb {\etal}
		\cite{Kolb:2005da}, who suggested that a related renormalization
		of the expansion rate \emph{at the present day}
		could be responsible for the presently observed accelerated
		expansion. This proposal was criticized on various grounds in Refs.
		\cite{Kolb:2005me,Geshnizjani:2005ce,Flanagan:2005dk,Hirata:2005ei,
		Notari:2005xk,Rasanen:2005zy,Moffat:2005zx,Brandenberger:2002sk,
		Brandenberger:2004ki,Martineau:2005aa,Martineau:2005zu}
		and elsewhere. The present discussion differs from that given by
		Kolb {\etal} because there is no suggestion here that
		renormalization of the acceleration rate would have experimental
		consequences for observers within any given horizon volume;
		indeed, precisely the \emph{opposite} is the case.
		The renormalization is only relevant for so-called ``meta-observers''
		who can simultaneously compare rates between many different
		Hubble patches; it is relevant for what we can compute,
		but not necessarily for what we can observe.}
	However, the presence of $\delta N_0$ does not affect
	$n$-point expectation values of $\zeta$ for $n > 1$, because
	it leads to disconnected contributions. These are not the
	contributions we wish to measure from CMB observations
	and therefore can be discarded in the present context.
		
	\subsection{The power spectrum of the primordial curvature perturbation}
	\label{sec:zeta-power-spectrum}
	
	The simplest possible graph corresponds to Eq.~\eref{eq:zeta-tree-spectrum},
	which arises from the product of the two
	linear factors in~\eref{eq:deltaN},
	\begin{equation}
		\parbox[t]{15mm}{
			\begin{fmfgraph}(40,6)
				\deltaNdefs
				\fmfpen{0.6thin}
				\fmfleft{l}
				\fmfright{r}
				\fmf{powerspectrum_plain}{l,r}
			\end{fmfgraph}
		}
		\mapsto \hspace{3mm}
		\left( \frac{\partial N}{\partial \phi_\ast} \right)^2
		\langle \delta\phi(\vect{k}_1) \delta\phi(\vect{k}_2) \rangle_\ast ,
		\label{eq:tree}
	\end{equation}
	where the subscript `$\ast$' denotes evaluation on the initial slice.
	Although Eq.~\eref{eq:tree} is clearly a tree diagram in terms of the
	$\delta N$ diagrammatic rules described in
	\S\ref{sec:renormalized-delta-n}, it
	implicitly contains loop corrections which are present in the
	initial correlator $\langle \delta\phi(\vect{k}_1) \delta\phi(\vect{k}_2)
	\rangle$.
	This illustrates a general feature which arises in calculating $\zeta$
	correlators using $\delta N$; there are two distinct types of loop.
	The first sort are the quantum-mechanical loops (called q-loops
	in Ref.~\cite{Lyth:2006qz}) present in the expectation values of the
	$\delta\phi$. The second sort are loops which arise from expanding
	$\zeta$ correlators using the $\delta N$ algorithm (called c-loops
	in Ref.~\cite{Lyth:2006qz}). To arrive at a consistent result at any
	given order in the loop expansion, it is necessary to retain loop
	contributions from \emph{both} these sources.
	Therefore, we require
	the q-loop correction in $\langle \delta\phi(\vect{k}_1)
	\delta\phi(\vect{k}_2) \rangle_\ast$.
	This q-loop has been computed by Sloth
	\cite{Sloth:2006az,Sloth:2006nu} in the limit where $t_\ast$ is long
	after the mode with wavenumber $k \sim k_1 = k_2$ exited the horizon,
	and has recently been given in Ref.~\cite{Seery:2007we}
	in the opposite case where $t_\ast$ is the horizon-crossing time
	associated with $k$. We are free to base the $\delta N$
	expansion on either of these slices, and a discussion of their
	merits is deferred to \S\ref{sec:slice} below.

	At whatever time it is evaluated,
	the q-loop associated with Eq.~\eref{eq:tree}
	is accompanied by a number of c-loops arising from
	$\delta N$ integrations.
	The first of these can be written
	\begin{equation}
		\parbox[t]{15mm}{
			\begin{fmfgraph}(40,6)
				\deltaNdefs
				\fmfpen{0.6thin}
				\fmfleft{l}
				\fmfright{r}
				\fmf{plain,tension=3}{l,v1}
				\fmf{plain,left}{v1,v2}
				\fmf{plain,right}{v1,v3}
				\fmf{plain,tension=4}{v4,r}
				\fmfpoly{shaded}{v2,v3,v4}
			\end{fmfgraph}
		}
		\mapsto
		(2\pi)^3 \delta(\vect{k}_1 + \vect{k}_2)
		\frac{\partial N}{\partial \phi_\ast}
		\frac{\partial^2 N}{\partial \phi_\ast^2}
		\int \frac{\d^3 q}{(2\pi)^3}
		B_\ast(\vect{k}_1,\vect{k}_2-\vect{q},\vect{q}) ,
		\label{eq:oneloop-bispectrum}
	\end{equation}
	where $B$ is the bispectrum of the field fluctuations,
	defined in Eq.~\eref{eq:threepoint},
	and the subscript `$\ast$' again denotes evaluation on
	the initial slice.
	This term gives a contribution to the loop correction which depends on
	the presence of non-gaussianity in the field perturbations $\delta\phi$.
	We shall see
	in \S\ref{sec:constant-tilt} below that this term can in fact give the
	largest contribution to the loop correction if $\fnl$ grows
	on very large scales, although this does not happen in the usual scenario.
	The presence of a single unconstrained momentum integral,
	here labelled $\vect{q}$, indicates that this is a
	one-loop contribution. The finite time of evaluation,
	at conformal time $\eta_\ast$, provides a
	natural upper cut-off in the momentum integral.
	On scales characterized by wavenumbers \emph{larger} than
	$q \sim -\eta_\ast^{-1}$
	no perturbations have yet been generated; on these scales, the
	spectrum and all higher correlation functions such as $B$ and
	the trispectrum $T$ are zero.%
		\footnote{If we were computing these objects using
		the underlying quantum field theory of $\zeta$ rather than
		by following the $\delta N$ method, then these wavenumbers
		would lie in a region of $\eta$-integration where the
		integrand is strongly suppressed by decaying exponentials,
		leaving almost no net contribution.}
	The upper limit on the integral therefore depends on when
	we take the initial slice to be constructed, whereas
	the comoving box size $\ell$ provides a natural
	(and time-independent) lower cutoff
	at $q \sim \ell^{-1}$.
	
	The next contribution arises from the product of two tree-level
	spectra. There are two ways such a product can arise. The
	first corresponds to the diagram
	\begin{equation}
		\parbox[t]{15mm}{
			\begin{fmfgraph}(40,6)
				\deltaNdefs
				\fmfpen{0.6thin}
				\fmfleft{l}
				\fmfright{r}
				\fmf{plain,tension=3}{l,v1}
				\fmf{powerspectrum_plain,left}{v1,v2}
				\fmf{powerspectrum_plain,right}{v1,v2}
				\fmf{plain,tension=3}{v2,r}
			\end{fmfgraph}
		}
		\mapsto
		(2\pi)^3 \delta(\vect{k}_1 + \vect{k}_2)
		\frac{1}{2}
		\left( \frac{\partial^2 N}{\partial\phi_\ast^2} \right)
		\int \frac{\d^3 q}{(2\pi)^3}
		P_\ast(|\vect{k}_1-\vect{q}|)P_\ast(r) ,
		\label{eq:oneloop-power-a}
	\end{equation}
	whereas
	the second corresponds to a different configuration
	\begin{equation}
		\parbox[t]{15mm}{
			\begin{fmfgraph}(40,6)
				\deltaNdefs
				\fmfpen{0.6thin}
				\fmfleft{l}
				\fmfright{r}
				\fmf{powerspectrum_plain}{l,v1}
				\fmf{powerspectrum_plain,left}{v1,v1}
				\fmf{plain}{v1,r}
			\end{fmfgraph}
		}
		\mapsto
		(2\pi)^3 \delta(\vect{k}_1 + \vect{k}_2)
		\frac{\partial N}{\partial \phi_\ast}
		\frac{\partial^3 N}{\partial \phi_\ast^3}
		\int \frac{\d^3 q}{(2\pi)^3}
		P_\ast(k)P_\ast(q) ,
		\label{eq:oneloop-power-b}
	\end{equation}
	and the remarks below Eq.~\eref{eq:oneloop-bispectrum}
	concerning the presence of an upper- and lower-cutoff in the
	momentum integral apply equally to both these loops.
	
	These diagrams exhaust the possible one-loop contributions if we
	assume that the one-point function of the field fluctuations is
	zero, $\langle \delta\phi(\vect{k}) \rangle = 0$. As discussed in
	\S\ref{sec:spectra}, the one-point function $O$ is given to one-loop
	order by
	a renormalization-scheme dependent number which can be absorbed into
	the background field $\phi(t)$. This number is accompanied by other
	unknown renormalization-scheme dependent numbers which are present
	in the two- and higher $n$-point correlation functions.
	We will assume that to a good approximation,
	all such unknown quantities can be ignored; since we will not keep
	track of scheme-dependent numbers in the higher $n$-point functions
	there is no point keeping track of $O$ or a renormalization of the
	background field $\phi(t)$ either.
	Therefore, all contributions
	from one-point insertions will be discounted
	in the remainder of this paper.

	These expressions agree with those recently given
	by Byrnes {\etal} in Ref.~\cite{Byrnes:2007tm}.
	
	\subsection{Where should the initial slice be set?}
	\label{sec:slice}
	
	In the discussion of \S\ref{sec:zeta-power-spectrum} the initial time
	at which the $\delta N$ expansion is to be constructed, $\eta_\ast$,
	was left arbitrary. The initial time plays two roles in the analysis:
	it determines whether the field correlators
	$\langle \delta \phi(\vect{k}_1) \cdots \delta \phi(\vect{k}_n)
	\rangle_\ast$ are to be evaluated at the approximate time
	when the mode with wavenumber $k \sim k_1 = k_2$ crosses the horizon,
	or if they are to be evaluated at a much later time;
	and it provides an upper cut-off in integrals over
	c-loop momenta. These two effects must combine in such a way that
	the curvature perturbation is independent of $\eta_\ast$.
	
	The loop correction evaluated long after horizon crossing has been
	computed by Sloth \cite{Sloth:2006az,Sloth:2006nu}
	in terms of the variance $\langle \delta\phi^2 \rangle \equiv
	\int \d \ln q \, \Ps(q)$, which is in principle a infrared
	divergent quantity
	and should be cut off at a momentum scale of order the box size,
	$q \sim \ell$, if $\Ps$ does not vanish as $q \rightarrow 0$.
	The loop correction is found to diverge
	essentially logarithmically with the time of observation,
	\begin{equation}
		\fl
		\langle \delta \phi(\vect{k}_1) \delta \phi(\vect{k}_2) \rangle_\ast
		\sim (2\pi)^3 \delta(\vect{k}_1 + \vect{k}_2)
		P_\ast(k) \Big\{
			1 + \alpha \Ci(-2k\eta_\ast)
			\langle \delta\phi^2 \rangle
		\Big\}
		\label{eq:sloth-loop}
	\end{equation}
	where $|k\eta_\ast| \ll 1$ by assumption,
	$\alpha$ is a numerical factor of order $\Or(\epsilon)$,
	$P_\ast$ is the tree-level power spectrum,
	which is assumed to be almost constant between horizon crossing
	and the time of observation, and $\Ci(x)$ is the real part of the
	imaginary exponential integral,
	\begin{equation}
		\fl
		\Ci(x) \equiv \Re \Big\{ \Ei_1(\im x) \Big\} ,
		\quad \mbox{where} \quad
		\Ei_n(z) \equiv \int_{1}^\infty
		  \frac{\d t}{t^n} \e{-t z} \sim
		- \gamma - \im \pi - \ln z
		\;\; \mbox{for $z \ll 1$} ,
		\label{eq:sloth-loop}
	\end{equation}
	where $\gamma$ is Euler's constant.
	Eq.~\eref{eq:sloth-loop} apparently contains \emph{two} distinct
	divergences.
	
	One divergence is implicit in $\langle \delta\phi^2 \rangle$.
	In exact de Sitter space $\Ps(q)$ is a constant, so
	$\langle \delta\phi^2 \rangle$ appears to contain a logarithmic
	divergence at zero momentum. This is caused a pile-up
	of modes outside the horizon, arising from gravitational
	redshifting, and is ultimately traceable to the quasi-de Sitter
	expansion. As modes are redshifted outside the horizon they carry
	perturbations, and the ever-increasing number of modes in this
	infrared phase space leads to large effects. However, the logarithmic
	divergence itself is fictional because the original inflationary
	patch was presumably not infinite in extent. Working in a finite box
	cuts off the divergence on a momentum scale $q \sim \ell^{-1}$.

	The second
	divergence is present in the $\eta_\ast \rightarrow 0$ limit,
	and becomes manifest when one observes the correlator
	long after the $\vect{k}$-mode in question has passed outside the
	horizon. This divergence comes from the logarithmic term in
	$\Ci(-2k\eta_\ast)$. As with the pile-up of modes, this divergence is
	fictional in the sense that no observer within the universe can
	really set $\eta_\ast = 0$. Instead, observations must be made at a finite
	time and this finite time gives a cut-off on $\eta_\ast$.

	Although we do not encounter a genuine infinity in either case,
	these ``divergences'' can generate large coefficients
	in the perturbation series.
	Consider first the divergence as $\eta_\ast \rightarrow 0$, which apparently
	signals a breakdown of perturbation theory at some time of order
	$N \sim 1/\epsilon$ e-folds after horizon crossing.
	(Related breakdowns have been reported elsewhere in the literature;
	see, for example, Refs.~\cite{Losic:2005vg,Wu:2006ew,Losic:2006ht}.)
	The presence of this term endows the correlator with a time dependence
	that is not visible at tree-level.
	Therefore one should worry whether this time dependence could spoil
	conservation of $\zeta$ on superhorizon scales.
	If so this would be cause for serious concern,
	because $\zeta$-conservation is known to hold to all orders
	of the $\delta N$ formula \cite{Lyth:2004gb}
	and therefore cannot be spoiled by c-loops, and since
	no field correlator is ever evaluated at the time $\zeta$ is observed
	it cannot be spoiled by q-loops among the $\{ \delta \phi \}$ either.
	
	In fact, the time dependence in
	Eq.~\eref{eq:sloth-loop} is merely what is \emph{required} to maintain
	conservation of $\zeta$. Consider any two widely separated initial
	times. When evaluated at the same time of observation,
	the coefficient $\partial N/\partial \phi_\ast$
	evolves between these two initial slices
	because the classical field $\phi$ is rolling down its
	potential. The correlator $\langle \delta\phi(\vect{k}_1)
	\delta\phi(\vect{k}_2) \rangle$ must likewise evolve between the
	two initial slices to compensate, in order
	that $\zeta$ remains constant. It is this evolution which is represented
	by the logarithmic ``divergence'' in~\eref{eq:sloth-loop}.
	Indeed, Eq.~\eref{eq:sloth-loop} should be thought of
	as the beginning of a Taylor series in $\ln(-k\eta_\ast)$,
	with the higher-order terms in the series generated by higher-order
	loop corrections.
	This series is of the form studied by Gong \& Stewart
	\cite{Gong:2001he,Gong:2002cx}. In particular,
	it implies that perturbation theory in powers of slow-roll parameters
	breaks down at sufficiently late times, of order $N \sim 1/\epsilon$
	e-folds after horizon crossing.
	
	Despite this breakdown in perturbation theory, our ability to make
	predictions from theories of the early universe is not really
	impaired. In such situations, one would ordinarily attempt to resum the
	large logarithms using a form of the renormalization group equation.
	This resummation procedure effectively moves the large coefficients
	from higher-order terms in the perturbation series into the
	leading-order term. In the context of Eq.~\eref{eq:sloth-loop},
	resummation of the large logarithms clearly
	yields the superhorizon evolution
	of the correlator. Since the perturbations behave classically in the
	superhorizon limit, we should expect that this evolution will
	correspond to the classical evolution of the field
	\cite{Starobinsky:1986fx,Starobinsky:1994bd,Tsamis:2005hd,Miao:2006pn}.

	Now consider the alternative choice, where the loop correction associated
	with a mode of wavenumber
	$\vect{k}$ is evaluated at the time when $\vect{k}$ crosses the horizon.
	At horizon crossing it follows that $-k\eta_\ast \simeq 1$.
	The large logarithmic term $\Ci(-2k\eta_\ast)$, and any higher-order
	logarithms, are therefore completely negligible.
	This version of the loop correction has recently
	been computed in Ref.~\cite{Seery:2007we}, with the result
	\begin{equation}
		\fl
		\langle \delta\phi (\vect{k}_1) \delta\phi (\vect{k}_2)
		\rangle_\ast =
		(2\pi)^3 \delta(\vect{k}_1 + \vect{k}_2)
		P_\ast(k) \left( 1 - \frac{4}{3}
			\Ps_\ast \ln k + \cdots
		\right) .
		\label{eq:oneloop-tree}
	\end{equation}
	The scale which accompanies $k$ inside the logarithm is set by the
	details of physics in the ultraviolet.%
		\footnote{In \texttt{v1}--\texttt{v2} of the \texttt{arXiv}
		version of this paper, and the version which subsequently appeared
		in JCAP, the regulating scale which accompanies the logarithm
		was erroneously identified as the infrared scale $\ell$.
		I would like to thank E. Dimastrogiovanni
		for drawing my attention to this error.
		In these versions the coefficient which accompanies $\ln k$
		was also given incorrectly, owing to a calculational error
		which was present in \texttt{v1}--\texttt{v2} of
		Ref.~\cite{Seery:2007we}. This error was corrected from
		\texttt{v3} onwards of that reference, to which the reader
		should refer for details.}
	Although such corrections are extremely interesting, they are not the
	focus of the present paper. For this reason we will ignore them
	in what follows.
	At this order, $\langle \delta\phi (\vect{k}_1) \delta\phi (\vect{k}_2)
	\rangle_\ast$ is free of infrared divergences.
	
	The key lesson to draw from a comparison of
	Eqs.~\eref{eq:sloth-loop}
	and~\eref{eq:oneloop-tree} is that
	the divergence at late times, associated with powers of
	$\ln(-k\eta_\ast)$, has been
	resummed into the $\delta N$ coefficient
	associated with this correlation function.%
		\footnote{There is nothing to prevent
		powers of $\ln(-k \eta_\ast)$ accompanying q-loop corrections
		such as Eq.~\eref{eq:oneloop-tree}. These are genuine quantum
		corrections to the time evolution, and we should not expect
		the classical $\delta N$ formula to account for their resummation.
		However, since these terms are suppressed by powers of
		$H^2$ we assume that their contribution is small.}
	It is a remarkable virtue of the $\delta N$ formulation that this
	resummation is
	implicit in our freedom to fix the initial slice at any convenient time,
	without any necessity to deploy a complex formalism based on the
	renormalization group.
	Eqs.~\eref{eq:sloth-loop} and~\eref{eq:oneloop-tree} are therefore
	merely different representations of the same physics. It does not matter
	which time of evaluation we pick, as long as we have the ability to
	perform the relevant calculations consistently.
	
	In this paper, we take the view that for analytic calculations
	it is better to evaluate the
	correlators of the fields soon after horizon crossing,
	where the logarithmic term $\Ci(-2k\eta_\ast)$, and any higher-order
	logarithms, are negligible. There are two key reasons to adopt
	this approach:
	\begin{enumerate}
		\item The q-loop correlators given in Eqs.~\eref{eq:sloth-loop}
		and~\eref{eq:oneloop-tree} were derived on the assumption that
		slow-roll is a good approximation between the time of horizon
		crossing and the time of observation of the $\delta\phi$.
		At any given order in
		the loop expansion one encounters a potentially large number of
		diagrams, but many of these diagrams will be subdominant
		when slow-roll applies.
		For example, as discussed in Ref.~\cite{Seery:2007we},
		the q-loop correction to the two-point function
		has a diagrammatic expansion of the form
		\begin{center}
			loop correction $\supseteq$ \hspace{1mm}
			\parbox{20mm}{\begin{fmfgraph*}(60,40)
				\fmfleft{l}
				\fmfright{r}
				\fmf{plain}{l,v}
				\fmf{plain}{r,v}
				\fmf{plain}{v,v}
			\end{fmfgraph*}}
			\hspace{1mm} + \hspace{1mm}
			\parbox{20mm}{\begin{fmfgraph*}(60,40)
				\fmfleft{l}
				\fmfright{r}
				\fmf{plain,tension=2}{l,v1}
				\fmf{plain,tension=2}{r,v2}
				\fmf{plain,left}{v1,v2}
				\fmf{plain,left}{v2,v1}
			\end{fmfgraph*}}
			\hspace{1mm}
			+ $\cdots$
		\end{center}
		Both of these diagrams are one-loop and therefore of
		second order in cosmological perturbation theory, which is an
		expansion in powers of $\Ps_\ast \simeq 10^{-10}$. However,
		the second diagram is suppressed by an extra factor of order
		$\sim \epsilon$, since each three-point vertex carries a factor
		of $\dot{\phi}/H$ \cite{Seery:2005gb} whereas the four-point
		vertex carries no suppression \cite{Seery:2006vu}. On the other hand,
		large logarithms such as $\Ci(-2k\eta_\ast)$
		can compensate for the smallness of
		$\epsilon$. Where large logarithms are present the first
		diagram is dominated by the terms of \emph{subleading} order
		in slow-roll, since it is these terms which are accompanied
		by large logarithmic factors \cite{Sloth:2006az,Sloth:2006nu}.
		When terms such as $\Ci(-2k\eta_\ast)$ are large
		one must be wary that the
		second diagram (among other possible sources of subleading
		slow-roll terms) does not also begin to contribute.
		As the logarithms become increasingly large, further diagrams
		at two-loops and beyond may also become relevant.
		Therefore, in order to compute reliably with
		a slow-roll truncated q-loop such as Eq.~\eref{eq:oneloop-tree}
		one should pick the point of evaluation in such a way that
		large logarithmic terms do not appear.
		This means that the initial hypersurface on which we base the
		$\delta N$ expansion should be close to the horizon crossing
		hypersurface.
		The slow-roll expansion is
		then safe between these two hypersurfaces,
		and we can compute the q-loop to
		very good accuracy by assuming it is dominated by the leading
		slow-roll part of the first diagram.
		For the two-point function
		this difficulty is not severe, since for moderate values of
		$\ln (-k\eta_\ast)$ one can account for the
		contribution of the second diagram, together with the subleading
		part of the first. However, for higher-order
		correlation functions this may become a
		practical consideration.
		
		\item Eqs.~\eref{eq:sloth-loop} and~\eref{eq:oneloop-tree}
		are computed from integrals over $\eta$ whose integrands
		are rapidly oscillating when $|k\eta| \gg 1$. This kills
		any contribution from very early times, leaving the integral
		dominated by times where $|k\eta| \lesssim 1$.
		The pre-factors in Eqs.~\eref{eq:sloth-loop}
		and~\eref{eq:oneloop-tree} are derived by approximating slowly
		varying quantities such as $H$ and $\epsilon$ at the time of
		horizon crossing. This will be a good approximation if the
		correlator is observed just a few e-folds after horizon
		crossing but may be on the verge of breaking down
		if $\ln |k\eta_\ast|$ is large, since in general one would then
		expect $H$ and $\eta$ to evolve between horizon crossing and the
		time of observation. One can account for this evolution using
		the $\delta N$ expansion.
	\end{enumerate}
	
	In the remainder of this paper only the ``early'' initial slice
	is considered, where the correlator is evaluated just a few e-folds
	after horizon crossing.
	
	\section{One-loop renormalization of the $\zeta$ power spectrum}
	\label{sec:rg-power}

	\subsection{The loop correction}
	\label{sec:rg-log}
	We are now in a position to apply these general principles to the
	object of central interest, the power spectrum $\langle
	\zeta(\vect{k}_1) \zeta(\vect{k}_2) \rangle$.
	After collecting terms in
	Eqs.~\eref{eq:tree}--\eref{eq:oneloop-power-b}, the connected part
	of the power spectrum can be written
	\begin{equation}
		\fl
		\langle \zeta(\vect{k}_1) \zeta(\vect{k}_2) \rangle =
		(2\pi)^3 \delta(\vect{k}_1 + \vect{k}_2)
		\left\{
			\left( \frac{\partial N}{\partial \phi_\ast} \right)^2
			P_\ast(k)
			+ \int \frac{\d^3 q}{(2\pi)^3} \; \Sigma(\vect{k},\vect{q})
		\right\}
		\label{eq:power-spectrum}
	\end{equation}
	where $\Sigma$ is defined by
	\begin{equation}
		\fl
		\Sigma \equiv \frac{\partial N}{\partial \phi_\ast}
		\frac{\partial^2 N}{\partial \phi_\ast^2}
		B_\ast(k,|\vect{k}-\vect{q}|,q) +
		\frac{\partial N}{\partial \phi_\ast}
		\frac{\partial^3 N}{\partial \phi_\ast^3}
		P_\ast(k) P_\ast(q)
		+ \frac{1}{2} \left( \frac{\partial^2 N}{\partial\phi_\ast^2} \right)^2
		P_\ast(|\vect{k}-\vect{q}|)P_\ast(q) .
		\label{eq:Sigma}
	\end{equation}
	In this formula
	$\vect{k}$ can be taken to be either $\vect{k}_1$ or $\vect{k}_2$,
	since its orientation is immaterial, and the
	integration $\d^3 q$ is restricted to momenta between
	$q \sim k$ and $q \sim \ell^{-1}$. As discussed above,
	the time $\eta_\ast$ at which we construct the initial slice
	for the $\delta N$ expansion is taken to be the horizon-crossing
	time for $\vect{k}$.

	To estimate the magnitude of the loop correction
	in~\eref{eq:power-spectrum} we must be able to evaluate
	the $\vect{q}$ integral, but this is not trivial. In general,
	$B(k_1,k_2,k_3)$ is a complicated function of its arguments
	\cite{Seery:2005gb,Allen:2005ye} and the integral is difficult
	to evaluate exactly. We can obtain a reasonable estimate
	by applying the method of
	Boubekeur \& Lyth \cite{Boubekeur:2005fj}. In this approximation, one
	supposes that the integral is dominated by its infrared singularities.
	Summing over all singularities, one finds that
	\begin{equation}
		P_\zeta(k) = P_\ast(k) \left\{
			\left( \frac{\partial N}{\partial \phi_\ast} \right)^2
			+ \Pi \right\} ,
		\label{eq:powerspectrum-loop}
	\end{equation}
	where $P_\ast$ is the tree-level power spectrum evaluated at horizon
	crossing and $\Pi$ is defined by
	\begin{equation}
		\Pi \equiv
		\left\{ \frac{\partial N}{\partial \phi_\ast}
			\frac{\partial^3 N}{\partial \phi_\ast^3} +
			\left( \frac{\partial^2 N}{\partial \phi_\ast^2} \right)^2
		\right\} A(k\ell)
		- \frac{24}{5} \fnl \frac{\partial N}{\partial \phi_\ast}
		\frac{\partial^2 N}{\partial \phi_\ast^2} B(k\ell)
		\label{eq:Pi} ,
	\end{equation}
	in which, as mentioned above, we have dropped the ultraviolet
	loop correction of Eq.~\eref{eq:oneloop-tree}.
	The functions $A$ and $B$ are defined by
	\begin{equation}
		\fl
		A \equiv \langle \delta\phi^2 \rangle = \int \frac{\d^3 q}{(2\pi)^3}
		P_\ast(q)
		\quad \mbox{and} \quad
		B \equiv \langle \fnl \delta\phi^2 \rangle =
		\int \frac{\d^3 q}{(2\pi)^3} \fnl P_\ast(q) .
	\end{equation}
	The central issue in the remainder of this section, which will also be
	important for the background renormalization described in
	\S\ref{sec:constant-tilt}
	below, is how one should obtain estimates of $A$ and $B$.

	\subsection{The approximation of constant tilt}
	\label{sec:constant-tilt}
	First assume that the power spectrum and $\fnl$ can be approximated by
	Eq.~\eref{eq:running-approximations} with constant tilt parameters
	$n-1$ and $\nnl$.
	The $A$ and $B$ functions now satisfy
	\begin{equation}
		A(k\ell) = \Ps_\ast y_{n-1}(k\ell)
		\quad \mbox{and} \quad
		B(k\ell) = \Ps_\ast y_{\nnl + n - 1}(k\ell) .
		\label{eq:ab-tiltdef}
	\end{equation}
	where $y_r(x)$ is Lyth's $y$-function \cite{Lyth:2006gd}, defined by
	\begin{equation}
		y_r(x) \equiv \frac{1}{r} \left( 1 - | x |^{-r}
		\right) .
		\label{eq:y}
	\end{equation}
	The parameters $\fnl$, $\nnl$ and $n-1$ are evaluated at $k$.
	Eq.~\eref{eq:ab-tiltdef} is approximately valid provided that the
	box size is not too much larger than the lengthscale associated
	with the mode of interest, $k^{-1}$, so that the approximation of
	fixed tilt is reasonable throughout the box.
	
	The loop correction is of second order in cosmological perturbation theory,
	which is an expansion in powers of $\Ps$. However, the coefficients
	in Eq.~\eref{eq:Pi} generated by the $\delta N$ expansion enter at different
	orders in the slow-roll approximation.
	In particular, $(N')^2$ is of order
	$\epsilon^{-1}$ whereas $(N'')^2 + N' N'''$ is of order $1$
	and $\fnl N' N''$ is of order $\epsilon^{1/2}$.
	Which of these terms is dominant depends critically on the
	tilts $n-1$ and $\nnl$.
	\begin{enumerate}
		\item \textit{Positive tilt}.\label{case:i}
				This is the case where the spectrum \emph{decreases} on
				large scales, also known as a blue spectrum. It is not
				generically produced by single-field models of inflation.
				For positive $r$, the $y$-function defined in Eq.~\eref{eq:y}
				satisfies $y_r(x) \sim 1/r$. If the spectral and
				bispectral tilts
				$n-1$ and $\nnl$ are both positive, then the loop
				correction is very small.
		\item \textit{No tilt}.
				For $r = 0$, the $y$-function behaves like $y_r(x)
				\sim \ln|x|$. If $n-1 = \nnl = 0$, it follows that
				all terms in $\Pi$ behave like logarithms of $k\ell$.
				To estimate its
				magnitude, suppose that the loop correction constitutes no
				more than some fraction $f\%$ of the tree-level. It follows
				that
				\begin{equation}
					\label{eq:c-loop}
					\Pi = \Ps_\ast (\Delta N - N_\ast) \approx
					\Ps_\ast \Delta N \lesssim \frac{f}{100} ,
				\end{equation}
				where $\Delta N$ is the total number of e-foldings between the
				scale $\ell$ and the end of inflation, and $N_\ast \approx 60$
				measures when the presently observable universe left the
				horizon in terms of e-folds before inflation ended. The second
				approximate equality is valid if the loop correction is large,
				since this entails $\Delta N \gg N_\ast$.
				Eq.~\eref{eq:c-loop} can be interpreted as a bound on
				$\Delta N$ in order that the loop correction does not
				significantly modify the tree-level prediction. This gives
				$\Delta N \lesssim (10^2 f^{-1} \Ps_\ast)^{-1}$,
				or $\Delta N \lesssim 10^8$ for $f \sim 1\%$.
		\item \textit{Negative tilt}.
				This is the case where the spectrum increases on large
				scales, also known as a red spectrum. It is the generic
				type of spectrum produced by single-field models of inflation.
				For $r < 0$, the $y$-function has asymptotic form
				$y_r(x) \sim x^{|r|}/|r|$ when $x \gg 1$.
				In this regime, which term dominates
				depends whether $\nnl \gtrless 0$: the term involving
				$\fnl$ will be dominant if $\nnl$ is negative, meaning that
				the non-gaussianity grows on large scales.%
					\footnote{This is not the usual situation, since in
					single-field slow-roll inflation $\fnl \propto \epsilon$
					\cite{Maldacena:2002vr},
					and therefore the non-gaussianity decays on large scales.
					At the times when
					the fluctuation on large scales was imprinted, $\epsilon$
					was much closer to zero.}
				For the purposes of
				making an estimate, however, it does not matter which term
				is the dominant one since in either case $\Pi \sim
				(k\ell)^t$, where $t > 0$ is the dominant tilt,
				and the difference in pre-factor is not significant.
				It follows that
				\begin{equation}
					\Pi \sim \frac{\alpha}{t} \Ps_\ast \e{Nt} ,
					\label{eq:tilt-constraint}
				\end{equation}
				where $\alpha$ is a numerical constant which absorbs the
				pre-factor. Eq.~\eref{eq:tilt-constraint} gives an extremely
				powerful constraint on how strongly $\Ps$ and $\fnl$ can tilt,
				but unfortunately it is untrustworthy precisely in the
				interesting limit where $k\ell \gg 1$ or if the tilt is
				significant. As discussed in
				\S\ref{sec:deltaN}, in this limit the approximation of constant
				tilt overpredicts the spectrum. In reality the spectrum does
				not grow like a constant power, but only a power of a
				logarithm.
	\end{enumerate}
	Therefore we conclude that the significance of the loop correction
	depends on the number of e-folds of inflation between the fundamental
	scale, $\ell$, and the scale of interest, $k^{-1}$.
	Except where there is a significant negative tilt the loop correction
	is small, perhaps
	of order $1\%$ in a model with $10^{7}$ e-folds of inflation.
	However, to study the case of negative tilt
	a more sophisticated analysis is required.
	
	\subsection{Monomial potentials.}
	\label{sec:monomials}
	To obtain a better prediction for $k\ell \gg 1$, or where there is
	negative tilt,
	consider any local field theory model of inflation, with potential given by
	the monomial \cite{Linde:1993xx}
	\begin{equation}
		V(\phi) = g \phi^r
		\label{eq:monomial}
	\end{equation}
	where $g$ is a coupling constant and $r$ is usually taken to be an
	integer.
	For $r=2$ and small values of the field this corresponds to an
	Nflation-type scenario \cite{Dimopoulos:2005ac},
	in which the quadratic behaviour arises as an
	approximation to a periodic axion potential. For larger $r$
	or large fields such
	monomials fit into the general framework of so-called chaotic
	inflation, where $\phi$ is taken to be descending from above the Planck
	scale with a spectrum of field values which may cause some regions of
	the universe to inflate at the expense of other regions in which the
	field value is not suitable for inflation to occur.
	
	Inflation occurs for field
	values greater than $\phi_f \equiv r/\sqrt{2}$.
	However, one cannot contemplate arbitrarily large values for $\phi$.
	Local field theory presumably breaks down as an approximate description of
	nature at least when $V \approx 1$ (in Planck units), which occurs at a
	field value $\phiplanck \equiv g^{-1/r}$.
	This is certainly the largest value of the field for which the tools
	used in the present analysis make sense, but there may be other effects
	which imply that our approximations break down even before one reaches the
	Planck scale.
	For example, the `self-reproduction' phase of
	eternal inflation ends when $2 \delta\phi^2 \gtrsim 2\epsilon/3$
	\cite{Linde:1993xx}, which implies a
	maximum field value $\phi_{\eternal} \approx (2\pi^2 r^2 g^{-1})^{1/(r+2)}$.
	One can imagine either of these field values as natural
	energy scales at which to specify
	the masses and coupling constants of an effective field theory which
	approximates some ultraviolet completion of the Standard Model plus
	gravity.
	
	Let us restrict attention to a regime where local field theory is
	a good approximation. 
	Any two arbitrary field values $\phi_a$ and $\phi_b$
	separated by $\Delta N$ e-folds of inflation
	are related via the rule
	\begin{equation}
		\phi_a^2 = \phi_b^2 + 2r \Delta N .
		\label{eq:field-values}
	\end{equation}
	In order to match the fluctuation amplitude
	$\Ps_\zeta = (24.1 \pm 1.3) \times 10^{-10}$%
		\footnote{The quoted value is for WMAP3+SDSS, ignoring any possible
		tensor contribution \cite{Spergel:2006hy}.}
	observed in CMB experiments, the coupling must be tuned to
	precisely satisfy
	\begin{equation}
		g = \frac{12\pi^2 r^2 \Ps_\zeta}{\phi_\ast^{r+2}} ,
		\label{eq:coupling}
	\end{equation}
	where $\phi_\ast$ is the field value when presently observable scales were
	leaving the horizon. This typically occurs
	roughly $N_\ast \approx 60$
	e-folds before the end of inflation in order that the
	flatness and horizon problems of classical cosmology are resolved.
	Eqs.~\eref{eq:field-values}--\eref{eq:coupling} therefore imply a very
	small coupling constant, which is hard to understand at the Planck scale on
	the basis of our present knowledge of physics.

	Suppose that inflation begins at some field value $\phi_{\uv}$.
	After a few e-foldings of inflation, we can suppose that the
	flat FRW metric~\eref{eq:background-metric} applies within the inflating
	volume. One therefore sets the
	box size $\ell$ to correspond to the horizon size at this time. The field
	value when any smaller scale $q$ exits the horizon
	is related to the field value
	at the onset of inflation by~\eref{eq:field-values}, giving%
		\footnote{In fact, $\phi_{\uv}$ should be reduced by a small amount
		since we are assuming that $\ell$ does not correspond precisely
		to the inflationary patch when $\phi = \phi_{\uv}$ but to
		some slightly later time. However, if the field is in slow-roll
		at this time then $\phi$ will not roll far during the few
		e-folds which elapse.}
	\begin{equation}
		\phi^2(q) = \phi_{\uv}^2 - 2 r \ln q\ell .
	\end{equation}
	It follows that $\langle \delta\phi^2 \rangle$ can be written
	\cite{Afshordi:2000nr,Sloth:2006az}
	\begin{equation}
		A \equiv \langle \delta\phi^2 \rangle = \frac{r}{r+4} \Ps_\zeta
		\Bigg\{
			\left( \frac{\phi_{\uv}}{\phi_\ast} \right)^{r+4} - 1
		\Bigg\} \approx
		\frac{r}{r+4} \Ps_\zeta \left( \frac{\phi_{\uv}}{\phi_\ast}
		\right)^{r+4} ,
		\label{eq:a-monomial-full}
	\end{equation}
	where the second approximate equality applies in the limit $\phi_\ast \ll
	\phi_{\uv}$. If inflation lasts for a large total number of e-folds
	$N \gg r$, then Eqs.~\eref{eq:field-values}
	and~\eref{eq:a-monomial-full} imply
	\begin{equation}
		A = \frac{r}{r+4} \Ps_\zeta \left( \frac{N}{N_\ast}
		\right)^{2+r/2} .
		\label{eq:a-monomial}
	\end{equation}
	This scaling of $\langle \delta\phi^2 \rangle \equiv A$
	with the total number of e-folds, namely
	$A \propto N^{2+r/2}$, was given earlier
	by Sloth \cite{Sloth:2006nu}.
	A similar procedure allows us to estimate $B$, giving
	\begin{equation}
		B = \frac{r^3}{r+2} \frac{\Ps_\zeta}{\phi_\ast^2} \Bigg\{
			\left( \frac{\phi_{\uv}}{\phi_\ast} \right)^{r+2} - 1
		\Bigg\} \approx
		\frac{r^2}{2 N_\ast(r+2)} \Ps_\zeta \left(
			\frac{N}{N_\ast}
		\right)^{1 + r/2} .
		\label{eq:b-monomial}
	\end{equation}
	In making this estimate, we have approximated
	$\fnl \sim \epsilon$ at each
	scale. Although this is not strictly numerically accurate
	\cite{Maldacena:2002vr},
	it should be sufficient to capture the tilt of the non-gaussian
	fraction. Indeed,
	Eqs.~\eref{eq:a-monomial}--\eref{eq:b-monomial} show that $A$ grows
	faster than $B$
	by a power of $N$, and therefore that the non-gaussian component
	is negligible for a large total number of e-folds.
	This is in agreement with our expectation that the contribution
	to the loop correction from any intrinsic non-gaussianity among the
	$\delta \phi$ should be suppressed if $\fnl$ decays on large scales.
	In addition, since both $A$ and $B$ only grow as powers of $N$, these
	equations justify the statements made above that a constant tilt
	overpredicts $\langle \delta\phi^2 \rangle$ for a large range of e-folds.
	In what follows we assume that $N$ is sufficiently large that it is a
	good approximation to
	ignore the intrinsically non-gaussian contribution.
	
	Let us estimate how many e-folds can elapse before the loop
	correction becomes significant enough to spoil the tree-level prediction.
	The coefficient of the $A$-term
	in Eq.~\eref{eq:Pi} is $\Or(1)$ in slow-roll,%
		\footnote{Note that in a monomial model there is a special
		simplification, since $N''' = 0$.}
	\begin{equation}
		\frac{\partial N}{\partial \phi_\ast}
		\frac{\partial^3 N}{\partial \phi_\ast^3} +
		\left(\frac{\partial^2 N}{\partial \phi_\ast^2}\right)^2 =
		\frac{1}{r^2} .
	\end{equation}
	Thus, in order that the loop contributes
	at no more than some fraction $f$ of the tree-level, we must require
	$\Pi(\partial N/\partial \phi_\ast)^{-2} \lesssim f$, or
	\begin{equation}
		N \lesssim N_\ast^{(6+r)/(4+r)} \Ps_\zeta^{-2/(4+r)}
		\left\{ 2(4+r) f \right\}^{2/(4+r)} .
		\label{eq:max-deltan}
	\end{equation}
	This gives constraints on $N$ which are summarised in
	Table~\ref{table:constraints}.
	\begin{table}
		\begin{center}
		    \footnotesize
		    \renewcommand{\arraystretch}{1.3}
			\begin{tabular}{c|ccc|ccc} \hline\hline
				& \multicolumn{3}{c|}{$\bm{r=2}$ \textbf{model}} & \multicolumn{3}{c}{$\bm{r=4}$ \textbf{model}} \\
				\textsc{significance} & \textbf{maximum $N$} & $N_{\mathrm{P}}$ & $N_{\eternal}$
					& \textbf{maximum $N$} & $N_{\mathrm{P}}$ & $N_{\eternal}$ \\ \hline
				$\bm{0.1 \%}$ & \multicolumn{1}{c|}{$4.06 \times 10^4$}
					& \multirow{3}{*}{$1.28 \times 10^{10}$} & \multirow{3}{*}{$5.03 \times 10^5$}
					& \multicolumn{1}{c|}{$8.65 \times 10^3$}
					& \multirow{3}{*}{$6.31 \times 10^5$} & \multirow{3}{*}{$2.50 \times 10^4$} \\
				$\bm{1 \%}$ & \multicolumn{1}{c|}{$8.74 \times 10^4$} & &
					& \multicolumn{1}{c|}{$1.54 \times 10^4$} & & \\
				$\bm{10 \%}$ & \multicolumn{1}{c|}{$1.88 \times 10^5$} & &
					& \multicolumn{1}{c|}{$2.74 \times 10^4$} & & \\ \hline \hline
			\end{tabular}
		\end{center}
		\caption{Limits on the total number of e-folds, $N$, in
		monomial models with $r=2$ and $r=4$.
		$N_{\mathrm{P}}$
		measures the number of e-folds between the Planck scale
		and the time presently observable scales were leaving the horizon,
		and $N_{\eternal}$ is a similar measure, with the Planck scale
		replaced by the `self-reproduction' scale of eternal inflation,
		$\phi_{\eternal}$ (defined in the text).
		\label{table:constraints}}
	\end{table}
	Note that
	the bounds arising from the loop contribution are roughly compatible with
	the bounds derived by Wu {\etal} \cite{Wu:2006ew} on the basis that
	amplification of the quantum geometry should not spoil scale-invariance
	of the spectrum.
	
	The numerical limits on $N$ cannot be taken literally for
	\textsf{loop:tree} ratios as large as $10\%$, because
	the estimates for $g$ and $\langle \delta\phi^2 \rangle$ [given in
	Eqs.~\eref{eq:coupling} and~\eref{eq:a-monomial-full}] were derived
	assuming that the tree-level prediction dominated the power spectrum,
	and this is no longer true when the loop constitutes an appreciable
	fraction of the tree-level. Likewise, as the significance of the loop
	increases,
	contributions from two-loop corrections and beyond may become relevant.
	Nevertheless, these limits show clearly
	that if inflation is taken to persist for
	an exponentially long time, corresponding to initial conditions at the
	Planck scale or the `self-reproduction' scale of eternal inflation, then
	the loop correction will be large and cannot be ignored.
	On the other hand, if inflation is only taken to consist of sufficient
	e-folds to solve the horizon and flatness problems,
	$N \sim N_\ast \approx 60$, then the loop correction will be
	completely negligible. These conclusions are in qualitative agreement with
	the results of Sloth~\cite{Sloth:2006nu}, who
	found corrections of $\Or(1)$ for inflation beginning at the
	self-reproduction scale. The qualitative disagreement
	between Refs.~\cite{Sloth:2006az,Sloth:2006nu} and the present paper
	on the one hand
	and the conclusions of Martineau \& Brandenberger
	\cite{Martineau:2005aa} on the other
	arises because in Ref.~\cite{Martineau:2005aa}
	the random distribution of phases present in the inflationary
	perturbations caused an overall cancellation of the effect.
	In the present formalism the distribution of phases is invisible;
	only the spectrum is relevant, not the modes themselves, and
	the large effect described in Table~\ref{table:constraints} is
	effectively the phase-space volume divergence given in Eq.~(46) of
	Martineau \& Brandenberger \cite{Martineau:2005aa}.
		
	\section{Discussion}
	\label{sec:conclude}
	\setcounter{footnote}{0}
	In this paper, the $\delta N$ formalism has been used to estimate the
	magnitude of the one-loop correction to the inflationary power spectrum in
	the case of a single-field, slow-roll model of the inflationary epoch. The
	calculation can be split into two separate parts. The first depends on a
	detailed estimate of the one-loop correction generated by interference
	among quantized field modes around the time of horizon crossing. This
	calculation has recently been performed in Ref.~\cite{Seery:2007we}. The
	second measures the pile-up of field modes outside the horizon which are
	generated by the quasi-de Sitter expansion. This accumulation of modes as
	inflation proceeds gives an ever-larger phase space for infrared
	divergences \cite{Martineau:2005aa},
	which is measured by the so-called c-loops of the $\delta N$
	formalism. In this paper these two parts are assembled, allowing an
	estimate of the magnitude of the loop correction which includes all
	relevant effects.
	
	The results is entirely consistent with our expectations
	from flat space quantum field theory:
	we are free to formulate the fundamental
	microscopic theory which governs inflation at any energy scale we wish,
	but if we choose to formulate it at an energy scale which is wildly
	different from that of the phenomena we wish to observe, then we must
	expect significant corrections to be generated by loop effects. Such
	corrections take the form of large logarithms that violate the predictions
	of na\"{\i}ve dimensional analysis. In the present case the loop correction
	is given by Eqs.~\eref{eq:powerspectrum-loop}--\eref{eq:Pi},
	the fundmental theory is formulated at a scale $\ell$, and we wish to
	predict phenomena associated with the CMB at a scale $k^{-1}$.
	The large logarithm which gives a significant loop correction is
	$N \sim \ln k\ell$.
	If $k \ell$ is exponentially large, which occurs in monomial
	models of inflation if initial conditions are set at around the Planck
	scale, then the loop correction has a sensitivity
	to the tilt of the field perturbation power spectrum.
	The $\delta N$ loop from the convolution of two power spectra
	is the most sensitive and scales
	like $N^{2 + n/2}$; the $\delta N$ loop from any intrinsic
	non-gaussianity among the $\delta \phi$ is suppressed, because $\fnl$
	scales in the opposite direction, giving
	$N^{1 + n/2}$.
	In the case of a negative tilt, which produces perturbations of increasing
	amplitude on \emph{large} scales, the loop correction rapidly becomes
	very significant. The correction is irrelevant only if inflation lasts for
	a small number of e-folds.
	
	The physical effect which demands that loop corrections be taken into
	account is nothing more than
	back reaction. Long ago, Salopek \& Bond
	\cite{Salopek:1990re} used the stochastic formalism to show that
	any initially homogeneous background field would develop wild fluctuations
	on ultra-large scales as inflation progressed. The large loop correction
	encountered in the limit of a large box is another manifestation of this
	effect. It measures the dispersion generated by comparing a correlator
	calculated in one region of the universe with the same
	correlator calculated in a different region, perhaps where the background
	fields have very different values. The correlator calculated in the
	large box is therefore not measureable in any experiment: it involves
	comparing the CMB temperature which would be seen in an ensemble of
	different universes, and no physical observer can make such a comparison.
	For this reason, the correlation is a ``computable'' in the sense of
	Witten \cite{Witten:2001kn}, but is not an observable.
	
	This result has no consequences for how we make or report cosmological
	observations, but has very significant consequences for how such
	observations are compared with theoretical models. In models where the
	fundamental theory is not taken to be valid within an exponentially large
	region, we are entitled to entirely disregard the question of loop
	corrections, since in this case they yield a negligible shift in
	tree-level values. This class of models includes the familiar single-field
	examples where inflation ends when the field reaches a certain point on
	its potential. In such cases we know the conditions which lead to the
	end of inflation and we always have the option to work within a small box.%
		\footnote{Indeed, in such a case, it would presumably make no difference
		even if we chose to work within the large box because the only
		distinction between widely separated locations would be a time
		delay. This can be gauged away by a reparametrization of time,
		meaning that the difference is not physical.
		This is the essentially the argument given by Unruh
		\cite{Unruh:1998ic}.}
	On the other hand, where isocurvature fields are present,
	these fields will generally acquire large fluctuations within the
	inflating region. If these fields play a role in ending inflation,
	then we may lose the option to compute within a small box, whereas the
	correlator within a large box may be afflicted with large loop
	corrections.
	This problem is perhaps most severe precisely for those models to which
	we presently attach the greatest
	interest -- those motivated from fundamental
	particle physics. For example, one popular class of such models arise
	from the so-called landscape of string vacua \cite{Giddings:2001yu}.
	In such models, one imagines
	the unwanted six dimensions predicted by string theory to be wrapped up
	on a Calabi--Yau three-fold which may be topologically quite non-trivial,
	as measured by the existence of a large number of homological cycles.
	These cycles can be wrapped by fluxes of $p$-form fields, giving rise to
	regions of the three-fold which have different values for the cosmological
	constant $\Lambda$, together with the other constants of nature
	\cite{Bousso:2000xa}, and in general many scalar fields will be present
	which measure the size and shape of the Calabi--Yau three-fold.
	These different regions
	may exhibit substantially different low-energy physics. In such
	models, the natural scale at which one computes the masses and couplings
	in the low-energy theory would be the Planck scale, or perhaps the
	self-reproduction scale of eternal inflation.
	Eqs.~\eref{eq:powerspectrum-loop}--\eref{eq:Pi} show that to compare any
	such theory to observation, one must account for large loop corrections
	which are inevitably encountered in passing from the Planck scale to
	the scale relevant for CMB predictions.
	
	This has some implications for how large-field models of inflation are
	compared with CMB observations.
	To understand how the Planck scale theory is related to the theory
	when scales associated with the CMB exited the horizon,
	one would need to integrate out the relevant degrees of freedom
	between $\Planck$ and the CMB scale. By analogy with conventional field
	theory, as one changes scale the effective lagrangian should
	be controlled by a flow equation analogous to Polchinksi's
	exact renormalization group equation \cite{Polchinski:1983gv}.
	In flat space theory this flow has well-understood effects, attracting the
	lagrangian to an invariant hypersurface on which non-renormalizable
	couplings are suppressed.	
	(This would refer to the effective field theory for $\zeta$, not to
	the couplings in the matter theory.)
	Unfortunately a complete understanding of this process in the cosmological
	case is not yet in hand, and it does not appear to be understood precisely
	how the Planck scale
	theory will be related to the theory in a small box.
	One possible approach is to use the method of Salopek \& Bond
	\cite{Salopek:1990re}, who gave a Fokker--Planck equation which determined
	the probability distribution of the classical fields on large scales.
	Having done so, it is possible to use a small box to determine the value of
	the correlation functions at each value of the allowed background fields,
	but one is then restricted to making probabilistic statements concerning
	what should be observed in the CMB. This issue deserves further study.
	
	In any case,
	considerable work remains to be done. The present analysis should be
	generalized to the case of multiple scalar fields in order that it
	can be applied to the best-motivated models from fundamental
	physics. This involves an extension of the q-loop calculation given
	in Ref.~\cite{Seery:2007we} to the case of many scalar fields.
	The calculation given by Weinberg~\cite{Weinberg:2005vy} shows that where
	$\zeta$ is accompanied by $\mathcal{N}$ isocurvature fields,
	we can expect the $\zeta$ loop to receive an enhancement by a
	factor $\sim \mathcal{N}$. However, in a multiple field scenario a
	more involved process is required to estimate $\langle
	\delta \phi^2 \rangle$ \cite{GarciaBellido:1995qq,Vernizzi:2006ve,
	Battefeld:2006sz,Yokoyama:2007uu},
	since the growth of $\Ps$ on superhorizon
	scales is no longer simple to predict. These issues should be
	addressed with some urgency.

	\ack
	I acknowledge support from PPARC under grant PPA/G/S/2003/00076.
	I would like to thank
	C. Byrnes, J. Lidsey, D. Lyth, K. Malik, A. Mazumdar,
	A. Riotto and M. Sloth for
	for useful conversations, and especially F. Vernizzi
	for lengthy conversations and correspondence which have helped
	clarify my understanding. I would like to acknowledge the hospitality
	of the Abdus Salam Institute for Theoretical Physics, Trieste,
	and the Department of Physics, University of Cardiff, where portions
	of the work outlined in this paper were carried out.
	I would like to thank the Yukawa Institute for Theoretical Physics, Kyoto,
	for their hospitality
	while participating in the programme
	\emph{Gravity and Cosmology 2007}, where this paper was completed.
	
	I would like to thank E. Dimastrogiovanni for drawing my attention
	to an error present in earlier versions of this paper
	(discussed below Eq.~\eref{eq:oneloop-tree}).
	
	\section*{References}
\providecommand{\href}[2]{#2}\begingroup\raggedright\endgroup

\end{fmffile}
\end{document}